\documentclass[floatfix,aps,amsmath,nofootinbib,twocolumn,10pt]{revtex4}

\usepackage{listings}
\usepackage{graphicx}
\usepackage{bm}
\usepackage{rotating}
\usepackage{array}
\usepackage{amsmath}
\usepackage{amssymb} 
\usepackage{mathrsfs} 
\usepackage{cancel}
\usepackage{diagbox}
\usepackage{color}

\def\({\left(}
\def\){\right)}
\def\[{\left[}
\def\]{\right]}

\def\e{\begin{equation}}
\def\q{\end{equation}}
\def\m{\begin{eqnarray}}
\def\n{\end{eqnarray}}
\newcommand{\Mov}[1]{{\color{black}{#1}}}

\begin{document}

\title{
Analytical Emulator for the Baryon \Mov{Density Distribution} inside the Fuzzy Dark Matter Soliton from Machine Learning} 

\author{Ke Wang$^{1}$}
\thanks{{wangke@lnnu.edu.cn}}
\author{Jianbo Lu$^{1}$}
\author{Man Ho Chan$^{2}$}
\affiliation{$^1$Department of Physics, Liaoning Normal University, Dalian 116029, China}
\affiliation{$^2$Department of Science and Environmental Studies, The Education University of Hong Kong, Hong Kong, China}

\date{\today}

\begin{abstract}
An empirical baryon density profile can be included in the Schrödinger-Poisson (SP) equations to influence the fuzzy dark matter (FDM) soliton formation. However, to probe the effects of baryon on the other dynamical evolutions of the FDM soliton, its equation of motion (EoM) inside the corresponding FDM soliton is needed.
In this paper, given an empirical baryon density profile, we first provide the cylindrical symmetric FDM soliton solution about the FDM density and the total potential of FDM and baryon. Then, \Mov{instead of EoM}, we build an \Mov{analytical emulator (AE) for the baryon density distribution} from the obtained FDM density and total potential by machine learning. Finally, we check that this \Mov{AE} works as well as an empirical baryon density profile for the FDM soliton formation, with the fractional errors $\lesssim0.04$. It should also work \Mov{as a kind of baryon EoM} for some other simple FDM soliton evolutions.

\end{abstract}

\pacs{???}

\maketitle


\section{Introduction}
\label{sec:intro}

As a promising candidate for dark matter~\cite{Rubin:1982kyu,Davis:1985rj,Clowe:2006eq}, the ultralight scalar field with spin-$0$, extraordinarily light mass ($m\sim10^{-22}~\rm{eV}/c^2$) and de Broglie wavelength comparable to a few kpc, namely fuzzy dark matter (FDM)~\cite{Hu:2000ke}, can form an equilibrium configuration with size smaller than its de Broglie wavelength, the so-called FDM soliton~\cite{Guzman:2004wj}. The subsequent 
evolution of FDM solitons are usually simulated numerically, including the 
perturbation, the interference/collision and the tidal disruption/deformation of FDM solitons~\cite{Guzman:2004wj,Bernal:2006ci,Paredes:2015wga,Edwards:2018ccc,Munive-Villa:2022nsr,Tan:2024spp}, according to the coupled Schrödinger–Poisson (SP) system of equations. 
If FDM solitons keep stationary in a fixed background potential due to supermassive black holes~\cite{Davies:2019wgi}, an additional cylindrical~\cite{Bar:2019bqz} or ellipsoidal~\cite{Tan:2024dne} baryon density profile, the shooting method or the successive over-relaxation method can be used to compute the eigenvalues of equilibrium configurations.

However, how to deal with the dynamic evolution of FDM solitons in an evolving background potential?
In this case, we should enlarge the SP system with an additional equation of motion (EoM) to describe the evolving background potential.
For example, to simulate the FDM soliton collisions in the evolving background potential due to an simultaneously evolving baryon density profile, the baryon EoM inside the FDM soliton should be built according to the interaction between FDM and baryon and the thermodynamical properties of baryon in galaxies.
Although some general DM-baryon relations in galaxies have been found, such as a universal constant for DM-baryon interplay~\cite{Chan:2019ukj}, the relations between the total mass of DM and baryon in galaxies~\cite{Chan:2022fzq}, the mass discrepancy-acceleration relation~\cite{McGaugh:2004aw}, the radial acceleration relation~\cite{McGaugh:2016leg} and so on, these relations are too rough to be used to construct the baryon EoM inside the FDM soliton.

In this paper, \Mov{instead of building the baryon EoM inside the FDM soliton directly,} we propose a novel method to construct \Mov{an analytical emulator (AE) for the baryon density distribution} given a mock data of FDM and baryon density profile inside the FDM soliton.
Usually, it's very easy to reconstruct a function of one variable $f(x)$ according to data, such as reconstruction of equation of state of dark energy $w(z)$, Hubble parameter $H(z)$, luminosity distance $D_L(z)$ or cosmic reionization history $x_{\rm e}(z)$ by MCMC global fitting~\cite{Collett:2019hrr}, principal components analysis~\cite{Huang:2017huh}, Gaussian process~\cite{Shafieloo:2012ht,Seikel:2012uu,Yang:2024rsy}, artificial neural networks(ANNs)~\cite{Wang:2019vxv} or genetic algorithms (GAs)~\cite{Arjona:2019fwb,Nesseris:2012tt}.
As for the function of multiple variables $f(x,y,...)$, machine learning methods including GAs and ANNs are also the better choices for its reconstruction from data, such as the built of the analytical emulator for the linear matter power spectrum $P(k;\omega_{\rm b}, \omega_{\rm m})$ by GAs~\cite{Orjuela-Quintana:2022nnq,Orjuela-Quintana:2023uqb,Orjuela-Quintana:2024hha,Bartlett:2023cyr} and the gravitational wave surrogate models by ANNs~\cite{Setyawati:2019xzw,Khan:2020fso,Thomas:2022rmc}.
Since \Mov{AE for the baryon density distribution} is supposed to be a function of FDM density and total potential $\rho_{\rm b,AE}(\rho,\Phi)$, we will turn to GAs to build it too.

This paper is organized as follows. In Section~\ref{sec:soliton_profile}, we solve the SP system with an empirical baryon density profile background in cylindrical coordinates. In Section~\ref{sec:profile2equation}, we extract a mock data of baryon density profile and then construct \Mov{AE for the baryon density distribution} by GA. In Section~\ref{sec:check}, we check our baryon \Mov{AE} by including it in an enlarged SP system. Finally, a brief summary and discussion are provided in Section~\ref{sec:sum}.

\section{FDM soliton solution with an empirical baryon density profile}
\label{sec:soliton_profile}
Given an empirical baryon density profile background, FDM obeys the following SP system,
\begin{equation}
\label{eq:sp} 
\begin{cases}
\begin{aligned}
&i\hbar\frac{\partial \psi}{\partial t}=\left(-\frac{\hbar^2}{2m}\nabla^2+m\Phi\right)\psi , \\
&\nabla^2 \Phi=4\pi G |\psi|^2 + \rho_{\rm b}, 
\end{aligned}
\end{cases}
\end{equation}
where $m$ is the mass of the FDM particle, $\psi$ is its wavefunction, and $\Phi$ is the total gravitational potential, which is sourced by the FDM density $\rho=|\psi|^2$ and the baryon density $\rho_{\rm b}$. For the Milky Way, as~\cite{Bar:2019bqz,Launhardt:2002tx}, we only consider the main contributions: a nuclear stellar disk (NSD) and a spherical nuclear stellar cluster (NSC)
\begin{align}
\label{eq:emp}
\nonumber
\rho_{\rm b}(R,z)&=\rho_{\rm NSD}(R,z)+\rho_{\rm NSC}(r),\\
\nonumber
\rho_{\rm NSD}(R,z)&=\frac{\bar{\rho}_{\rm NSD}}{1+\left(\frac{R}{250{\rm pc}}\right)^{14}}\left[1-\tanh^4\left(\frac{R}{140{\rm pc}}\right)\right]e^{-\frac{|z|}{15{\rm pc}}},\\
\rho_{\rm NSC}(r)&=\frac{\bar{\rho}_{\rm NSC}}{1+\left(\frac{r}{0.22{\rm pc}}\right)^{n_{\rm NSC}}},~~r\leq200{\rm pc},
\end{align}
where $r=\sqrt{R^2+z^2}$, $\bar{\rho}_{\rm NSD}=330M_{\odot}/{\rm pc}^3$, $\bar{\rho}_{\rm NSC}=3.3\times10^6M_{\odot}/{\rm pc}^3$ and $n_{\rm NSC}=2$ for $r<6{\rm pc}$, $\bar{\rho}_{\rm NSC}=9.0\times10^7M_{\odot}/{\rm pc}^3$ and $n_{\rm NSC}=3$ for $r\geq6{\rm pc}$ and NSC has an outer cut-off radius at $r=200{\rm pc}$.

For this cylindrically symmetric baryon density profile, the waveform features an ansatz of $\psi(t,R,z) =e^{-i\gamma t/\hbar}\phi(R,z)$, where $\gamma$ is the ansatz energy eigenvalue.
Then the FDM soliton density
$\rho(R,z)=|\psi|^2=\phi^2(R,z)$ is simply related to the FDM soliton mass $M=\int_{R=0}^{R=\infty}\int_{z=-\infty}^{z=+\infty}2\pi R\rho(R,z)dRdz$, hence the total baryon mass inside the FDM soliton $M_{\rm b}=\int_{R=0}^{R=\infty}\int_{z=-\infty}^{z=+\infty}2\pi R\rho_{\rm b}(R,z)dRdz$.
Also with this cylindrical ansatz, Eq.~(\ref{eq:sp}) can be simplified to
\begin{equation}
\label{eq:sp1} 
\begin{cases}
\begin{aligned}
&\tilde{\nabla}^2\tilde{\phi}=2(\tilde{\Phi}+\tilde{\gamma})\tilde{\phi},\\
&\tilde{\nabla}^2\tilde{\Phi}=\tilde{\phi}^2+\tilde{\rho}_{\rm b},
\end{aligned}
\end{cases}
\end{equation}
where a set of dimensionless variables is defined as
\begin{align}
\nonumber
&\tilde{\phi}\equiv\frac{\hbar\sqrt{4 \pi G}}{mc^2} \phi,
~~
\tilde{\rho}_{\rm b}\equiv\left(\frac{\hbar\sqrt{4 \pi G}}{mc^2}\right)^2\rho_{\rm b},\\
\nonumber
&\tilde{M} \equiv \frac{GMm}{\hbar c},
~~
\tilde{M}_{\rm b} \equiv \frac{GM_{\rm b}m}{\hbar c},\\
&\tilde{r}\equiv\frac{mc}{\hbar}r,
~~
\tilde{\Phi}\equiv\frac{1}{c^2}\Phi,
~~
\tilde{\gamma}\equiv\frac{1}{mc^2}\gamma.
\end{align}
This dimensionless SP system obeys the scaling symmetry when
\begin{align}
\label{eq:symmetry}
\nonumber
&\tilde{\phi}  \longrightarrow  \lambda \tilde{\phi}, 
~~
\tilde{\rho}_{\rm b} \longrightarrow  \lambda^{2} \tilde{\rho}_{\rm b},\\
\nonumber
&\tilde{M} \longrightarrow  \lambda^{1/2} \tilde{M},
~~
\tilde{M}_{\rm b} \longrightarrow  \lambda^{1/2} \tilde{M}_{\rm b}\\
&\tilde{r}  \longrightarrow  \lambda^{-1/2} \tilde{r},
~~
\tilde{\Phi}  \longrightarrow  \lambda \tilde{\Phi}, 
~~
\tilde{\gamma}  \longrightarrow  \lambda \tilde{\gamma}. 
\end{align}

\begin{figure}[]
\begin{center}
\includegraphics[width= 9.4cm]{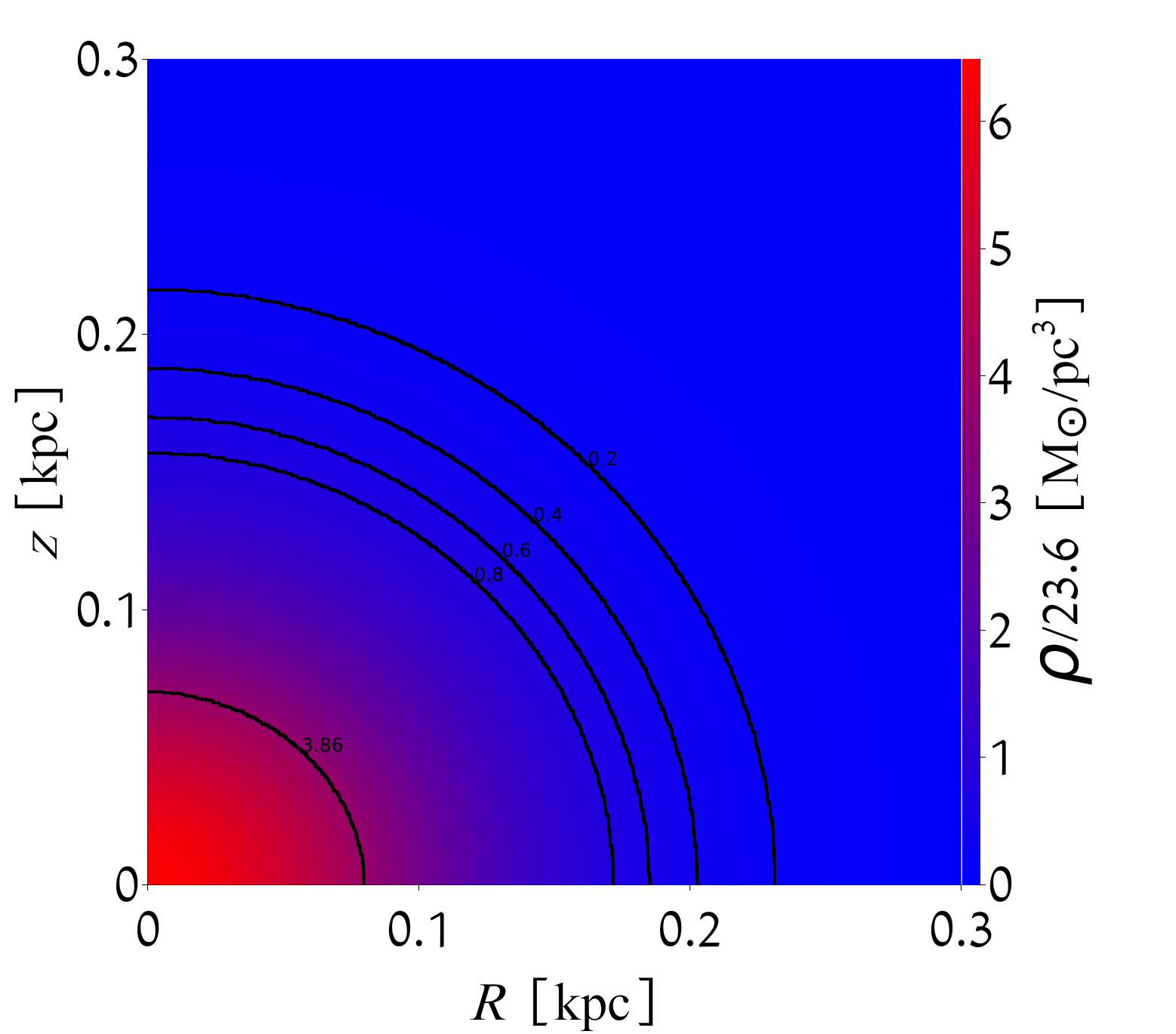}
\end{center}
\caption{FDM soliton density contours in the inner Milky Way, which is consistent with Fig.~2 of~\cite{Bar:2019bqz}. To compared with Fig.~2 of~\cite{Bar:2019bqz} easily, the density is also normalised to a reference value of $23.6M_{\odot}/{\rm pc}^3$ as Fig.~2 of~\cite{Bar:2019bqz}. The corresponding dimensionless total baryon mass is $\tilde{M}_{\rm b}=1.03$.}
\label{fig:rhoRz}
\end{figure}

\begin{figure}[]
\begin{center}
\includegraphics[width= 9.4cm]{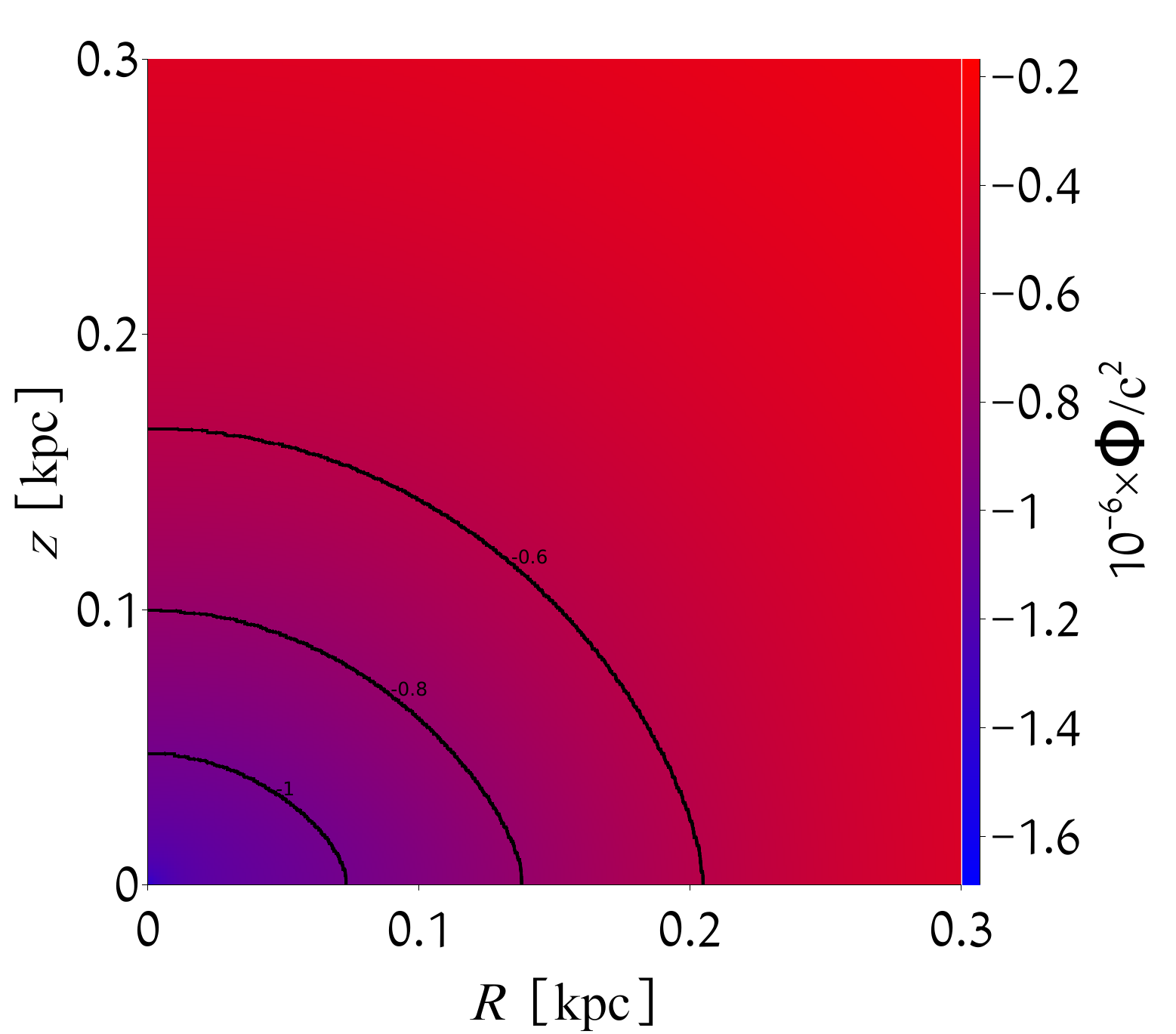}
\end{center}
\caption{Total potential contours inside the FDM soliton, where $\tilde{\Phi}(\tilde{R}=0,\tilde{z}=0)=-2.75395$ which is accompanied by $\tilde{\rho}_{\rm b}(\tilde{R}=0,\tilde{z}=0)=21551.2$ and $\tilde{\phi}^2(\tilde{R}=0,\tilde{z}=0)=1$.}
\label{fig:PhiRz}
\end{figure}

Given an initial $\lambda_0$, one can derive the corresponding $\tilde{\rho}_{\rm b}(\tilde{R},\tilde{z})$ from $\lambda_0^{2}\tilde{\rho}_{\rm b}({mc^2}/{\hbar\sqrt{4 \pi G}})^2=\rho_{\rm b}$ according to Eq.~(\ref{eq:emp}).
With this derived $\tilde{\rho}_{\rm b}(\tilde{R},\tilde{z})$, one can solve Eq.~(\ref{eq:sp1}) in cylindrical coordinates by following the Appendix~A of~\cite{Bar:2019bqz}.
When we confine ourselves to the normalized solution $\tilde{\phi}(\tilde{R}=0,\tilde{z}=0)=1$, a new $\lambda_1\neq1$ can be derived from $\lambda_1^{1/2}\tilde{M}\hbar c/Gm=M$, where $M$ can be predicted from the halo mass $M_{\rm{halo}}$ (for the Milky Way, $M_{\rm{halo}}=1\times 10^{12} ~M_{\odot}$~\cite{Wang:2019ubx}) according to the soliton-halo mass relation~\cite{Schive:2014hza}
\begin{equation} 
\label{eq:r_h}
M \approx 1.35\times 10^{9}\left(\frac{M_{\rm{halo}}}{10^{12}M_{\odot}}\right)^{1/3} \left(\frac{m}{10^{-22}~{\rm{eV}}/c^2}\right)^{-1}M_{\odot}.
\end{equation}
When $\lambda_0=\lambda_1$, it means that $\rho_{\rm b}$ and $M$ are equal to the baryon density and the FDM soliton mass of the Milky Way at the same time.
Therefore, if $\lambda_0\neq\lambda_1$, we replace $\lambda_0$ with $\lambda_1$ and repeat the above three steps~\cite{Tan:2024dne}. We stop the iteration until $\lambda_1/\lambda_0-1>0.0001$ or $\lambda_0\approx\lambda_1\approx6\times10^{-7}$ for\footnote{Here $m=10^{-22}~\rm{eV}/c^2$ just serve as an example for didactic purpose. Although different $m$ leads to different results in the following discussion, the method proposed in this paper could still hold.} $m=10^{-22}~\rm{eV}/c^2$.
In Fig.~\ref{fig:rhoRz} and Fig.~\ref{fig:PhiRz}, we show the final solutions of $\rho$ and $\Phi$, which are consistent with the results of ~\cite{Bar:2019bqz}. These solutions are dependent on $\tan\theta=\frac{z}{R}$, which will serve as mock data to construct the baryon \Mov{AE} inside the FDM soliton.

\section{Construction of baryon AE by machine learning}
\label{sec:profile2equation}
GA as a machine learning technique is inspired by the concepts in evolutionary biology: a population evolves generation by generation to adapt to the environment; a population is composed of many individuals; each individual is characterized by its own chromosome; several genes form the chromosome; every gene includes some nucleotides.
Therefore, making an analogy with the individual taking the form of
\begin{equation}
\label{eq:indiv}
{\rm individual}=\sum^{\rm chromosome}{\rm gene}({\rm nucleotides}),
\end{equation}
in this section, we look for a mathematical expression optimizing the goodness of fit to the given data of $\tilde{\rho}_{\rm b}$.
First, we describe how the mock data of $\tilde{\rho}_{\rm b}$ for the fitting process was gathered. Then we construct the fitting function or \Mov{AE for the baryon density distribution} by GA.

\subsection{Mock data}
Since Fig.~\ref{fig:rhoRz} shows the distribution of $\rho(R,z)$, from it and by setting $\lambda\approx6\times10^{-7}$, we can extract $\tilde{\phi}^2(\tilde{R},\theta)$ for a given pair of $\{\tilde{R},\theta\leq45^{\circ}\}$ or $\tilde{\phi}^2(\tilde{z},\theta)$ for a given pair of $\{\tilde{z},\theta>45^{\circ}\}$. Here, we consider that $\theta=0^{\circ},1^{\circ},...,90^{\circ}$ and $\Delta \tilde{R}=\Delta \tilde{z}=0.01$. However, we only sample $91\times91$ $\tilde{\phi}^2$ points by $91\times46$ pairs of $\{\tilde{R},\theta\leq45^{\circ}\}$ and $91\times45$ pairs of $\{\tilde{z},\theta>45^{\circ}\}$. 
Similarly, we also sample $91\times91$ $\tilde{\Phi}$ points by the same pairs of $\{\tilde{R},\theta\leq45^{\circ}\}$ and $\{\tilde{z},\theta>45^{\circ}\}$ from Fig.~\ref{fig:PhiRz}.
Combing the corresponding $91\times91$ $\tilde{\rho}_{\rm b}$ points retrieved from Eq.~(\ref{eq:emp}), our final dataset is a table of dimensions $8281\times4$ with data points given as $\{\theta,\tilde{\phi}^2,\tilde{\Phi},\tilde{\rho}_{\rm b}\}$.
We quantify the goodness of fit of a given analytical
expression for baryon \Mov{AE} by the following function
\begin{equation}
\%{\rm Acc}=\frac{100}{N}\sum_i^N\left|\frac{\lg[\tilde{\rho}_{{\rm b},i}]-\lg[\tilde{\rho}_{{\rm b},i,{\rm AE}}]}{\lg[\tilde{\rho}_{{\rm b},i}]}\right|
\end{equation}
where $N=91\times91=8281$ is the number of data points in our
dataset and $\tilde{\rho}_{{\rm b},i}$ is a simplified notation for $\tilde{\rho}_{\rm b}$ evaluated at $\{\theta_i,\tilde{\phi}^2_i,\tilde{\Phi}_i\}$.

\subsection{\Mov{AE for the baryon density distribution}}
In fact, construction of \Mov{AE for the baryon density distribution} belongs to a symbolic regression problem where both the model structure and parameters should be discovered from the data. In particular, the model structure can be fixed after a large number of generations by GA's genetic operators, namely, crossover and mutation.

GA looks for a fitting formula to a dataset by evolving the population, each individual of which takes a form similar to Eq.~(\ref{eq:indiv}). More precisely, in this paper, we set the analytical expression for baryon \Mov{AE} as
\begin{align} 
\label{eq:eom}
\nonumber
\lg[\tilde{\rho}_{\rm b,AE}(x,\theta)]&\equiv\lg21551.2 +f_{\rm AE}(3.75395,\theta)-f_{\rm AE}(x,\theta),\\
\nonumber
f_{\rm AE}(x,\theta)&\equiv\sum_{i=1}^7a_i(\theta)g_i\{b_i(\theta)[x-c_i(\theta)]\}d_i(x),\\
x&\equiv\tilde{\phi}^2-\tilde{\Phi},
\end{align}
where $21551.2$ comes from the maximum value of the mock data of $\tilde{\rho}_{\rm b}$, $3.75395$ is the value of $\tilde{\phi}^2-\tilde{\Phi}$ at $(\tilde{R}=0,\tilde{z}=0)$, functions $\{g_i(x)\}$ determine the model structure and $\{a_i,b_i,c_i,d_i\}$ serve as the ``amplitude'', ``frequency'', ``phase" and a correction respectively. The sum goes from 1 to 7 because the chromosome of each individual contains 7 genes and each gene is composed of several nucleotides $\{a_i,b_i,c_i,d_i,g_i\}$. $g_i$ is an operation in the grammar set $\{x,e^x,-e^{-x}\}$ and, for simplicity, we fix them as
\begin{align}
\label{eq:g}
\nonumber
g_1(x)&=g_2(x)=e^x,\\
\nonumber
g_3(x)&=g_4(x)=g_5(x)=g_6(x)=-e^{-x},\\
g_7(x)&=-x.
\end{align}
$\{a_i,b_i,c_i\}$ are linear functions of $\sin\theta$ to characterize the cylindrical symmetry of $f_{\rm AE}(x,\theta)$
\begin{align} 
\label{eq:theta}
\nonumber
a_i(\theta)&=a'_i(1+a''_i\sin\theta),\\
\nonumber
b_i(\theta)&=b'_i(1+b''_i\sin\theta),\\
c_i(\theta)&= c'_i(1+c''_i\sin\theta),
\end{align}
where $\{a'_i,b'_i,c'_i\}\in[0,5]$ and $\{a''_i,b''_i,c''_i\}\in[-1,1]$ are free parameters.
$d_i$ is a correction that can be fine-tuned to improve the goodness of fit
\begin{equation}
\label{eq:correct}
d_i(x)=\{1+0.18\left|j_1[1.5\pi(x-3.75395)]\right|\}(\delta_{1i}+\delta_{2i}),\\    
\end{equation}
where $\delta_{1i}$ and $\delta_{2i}$ mean that this correction only works for $i=1$ and $i=2$ cases and $j_1(x)=(\sin x-x\cos x)/x^2$ is the spherical Bessel function of order 1.

\begin{figure*}[]
\begin{center}
\includegraphics[width= 8cm]{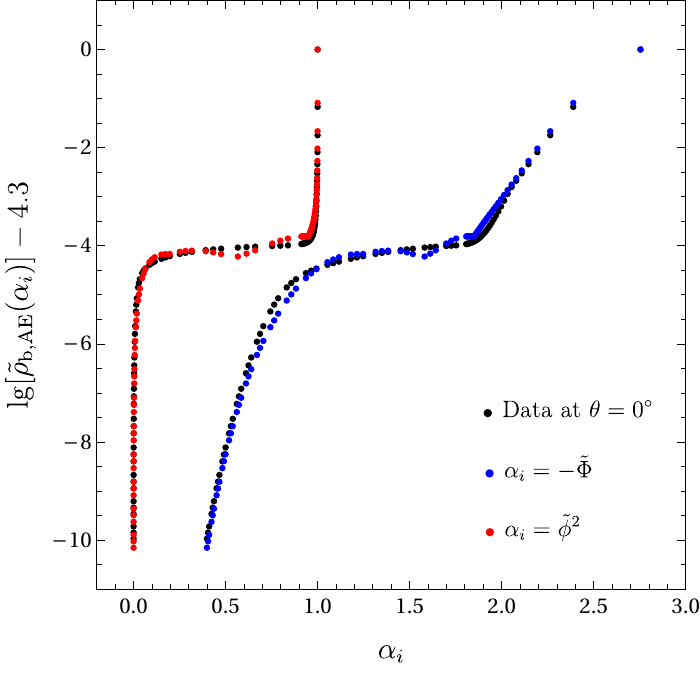}
\includegraphics[width= 8cm]{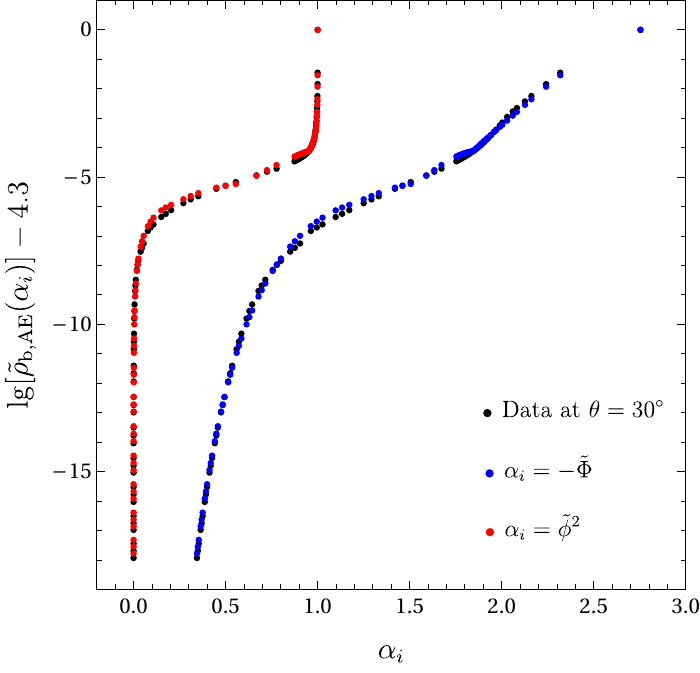}\\
\includegraphics[width= 8cm]{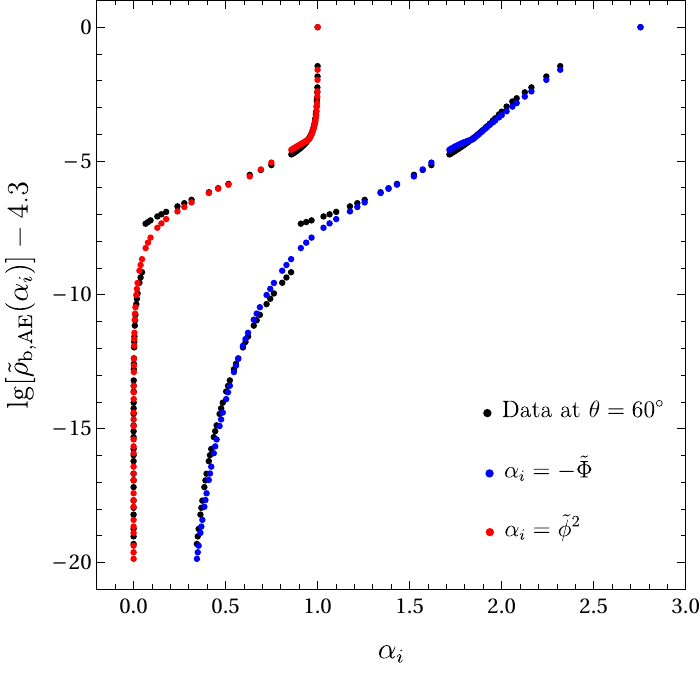}
\includegraphics[width= 8cm]{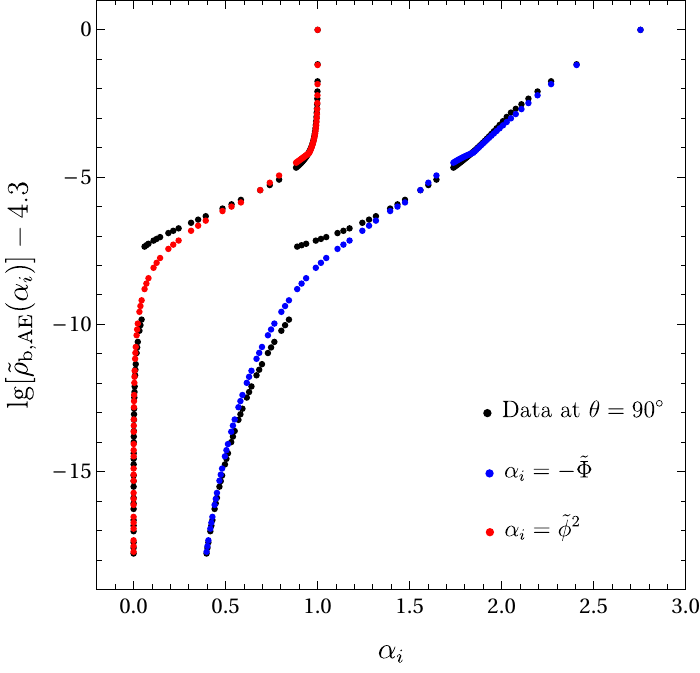}
\end{center}
\caption{Comparison between the baryon \Mov{AE} $\tilde{\rho}_{\rm b,AE}(x,\theta)$ (colored points) and the mock data $\tilde{\rho}_{\rm b}(x,\theta)$ (black points) at four angles $\theta=\{0^{\circ},30^{\circ},60^{\circ},90^{\circ}\}$ respectively. The unphysical jump in $\tilde{\rho}_{\rm b}(x,\theta)$ for larger $\theta$ comes from the cutoff of Eq.~(\ref{eq:emp}), which decreases the accuracy to some extent.}
\label{fig:anlge4}
\end{figure*}

As shown in Fig.~\ref{fig:anlge4}, each set of mock data (black points) includes three parts: the left part drops dramatically and behaves as $-e^{-x}$, the middle part features a kind of plateau of slow growth and behave as $x$, the right part increases dramatically and behaves as $e^x$. Furthermore, the unphysical jump for larger $\theta$ divides the left part into two sub parts. When we describe each part (or sub part) with the combination of at least two corresponding elementary functions, there would be $8$ $g_i(x)$. Since the combination of two linear functions can be replaced with another new linear function, we finally fixed $7$ $g_i(x)$ as Eq.~(\ref{eq:g}) before the code starts running.
Due to the cylindrical symmetry, the ``amplitude'', ``frequency'' and ``phase" $\{a_i,b_i,c_i\}$ should depends on $\theta$. When we choose $\theta=0^{\circ}$ as the pivot angle, the dependence of $\{a_i,b_i,c_i\}$ on $\theta$ at the other angles can be simply included as Eq.~(\ref{eq:theta}).
After many trials, we find that any combination of $g_1(x)$ and $g_2(x)$ can not fit the right part of mock data well. To improve the goodness of fit, we introduce corrections $\{d_i\}$ as Eq.~(\ref{eq:correct}) and fine tune their factors as $0.18$. 
In principle, GA can automatically discover the counterparts of Eq.~(\ref{eq:g}), Eq.~(\ref{eq:theta}) and Eq.~(\ref{eq:correct}) during the stochastic search. However, in practice, we provide them in advance to reduce the computational cost.

\begin{figure}[]
\begin{center}
\includegraphics[width= 8cm]{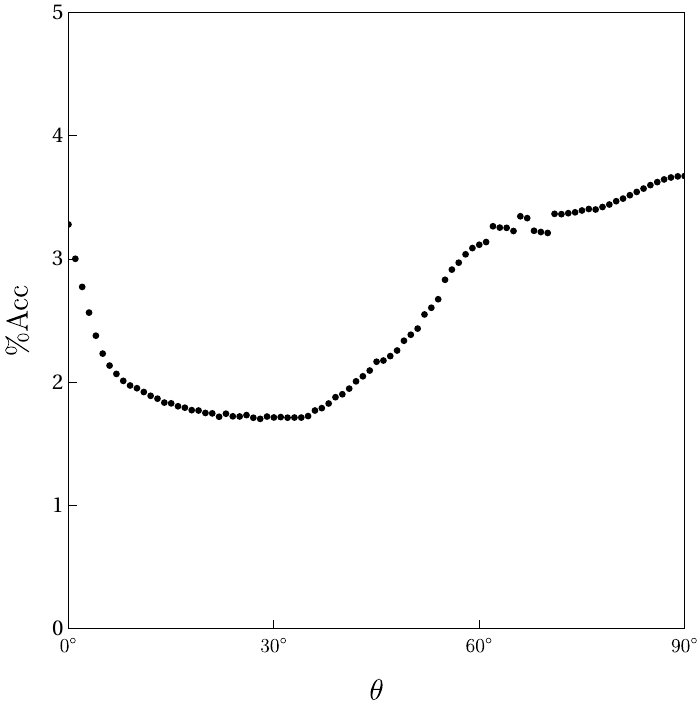}
\end{center}
\caption{Accuracy of the baryon \Mov{AE} as a function of $\theta$.}
\label{fig:acc}
\end{figure}

We perform a stochastic search with GA code~\cite{Arjona:2019fwb,Orjuela-Quintana:2022nnq,Orjuela-Quintana:2023uqb,Orjuela-Quintana:2024hha} and find a function with $\%{\rm Acc}=2.54844$, whose parameters are listed in Tab.~\ref{tb:EoM}. We also plot the accuracy of this baryon \Mov{AE} at different $\theta$ in Fig.~\ref{fig:acc}. The accuracy is $<4\%$, which means that our emulator can be used for some simple baryon evolutions inside the FDM soliton. In fact, the accuracy at $\theta\gtrsim60^{\circ}$ can be further improved if the unphysical jump in mock data $\tilde{\rho}_{\rm b}$ due to the cutoff of Eq.~(\ref{eq:emp}) disappears. In Fig.~\ref{fig:anlge4}, the comparison between $\tilde{\rho}_{\rm b,AE}(x,\theta)$ (colored points) and $\tilde{\rho}_{\rm b}(x,\theta)$ (black points) at four angles $\theta=\{0^{\circ},30^{\circ},60^{\circ},90^{\circ}\}$ is shown respectively. 

\begin{table}[htbp] 
\renewcommand\arraystretch{1.5}
\caption{Best-fit parameters for the baryon \Mov{AE} given in Eq.~({\ref{eq:eom}}).} 
\label{tb:EoM}
\begin{tabular}{|c| c c c c c c|} 
\hline 
 & $a'_i$ & $a''_i$ & $b'_i$ & $b''_i$ & $c'_i$ & $c''_i$\\ 
\hline 
$i=1$ & 1.35941 & -0.31828 & 0.75068 & -0.02252 & 0.45798 & 0.04361\\ 
$i=2$ & 1.11927 & 0.39288  & 0.19257 & -0.90871 & 2.99997 & -0.90871\\ 
$i=3$ & 1.11756 & 0.22907  & 0.56427 & -0.56092 & 2.92398 & -0.48616\\ 
$i=4$ & 3.83729 & 0.54773  & 0.89039 & -0.15683 & 0.73105 & -0.72978\\ 
$i=5$ & 0.05903 & 0.88515  & 3.49499 & 0.99812  & 1.37196 & -0.41290\\ 
$i=6$ & 0.31317 & 0.99983  & 3.88515 & -0.07250 & 1.01848 & 0.02457\\ 
$i=7$ & 2.00000 & -0.20860 & 2.88828 & -0.61984 & 1.52037 & -0.72819\\ 
\hline
\end{tabular}
\end{table}

\section{FDM soliton formation with the baryon AE}
\label{sec:check}
Note that the above baryon \Mov{AE} is constructed from dimensionless mock data $\{\theta,\tilde{\phi}^2,\tilde{\Phi},\tilde{\rho}_{\rm b}\}$. Therefore, replacing the fixed baryon background $\tilde{\rho}_{\rm b}$ of Eq.~(\ref{eq:sp1}) with an baryon \Mov{AE} $\tilde{\rho}_{\rm b,AE}$ (Eq.~(\ref{eq:eom})) leads to a new coupled system
\begin{equation}
\label{eq:sp2} 
\begin{cases}
\begin{aligned}
&\tilde{\nabla}^2\tilde{\phi}=2(\tilde{\Phi}+\tilde{\gamma})\tilde{\phi},\\
&\tilde{\nabla}^2\tilde{\Phi}=\tilde{\phi}^2+\tilde{\rho}_{\rm b,AE},\\
&\lg\tilde{\rho}_{\rm b,AE}=4.3 +f_{\rm AE}(3.75,\theta)-f_{\rm AE}(x,\theta),
\end{aligned}
\end{cases}
\end{equation}
where $x=\tilde{\phi}^2-\tilde{\Phi}$ means that this new coupled system does not satisfy the scaling symmetry as the original Eq.~(\ref{eq:sp1}) under transformation Eq.~(\ref{eq:symmetry}). However the dimensionless solutions of Eq.~(\ref{eq:sp2}) $\{\tilde{\phi}(\tilde{R},\tilde{z}),\tilde{\Phi}(\tilde{R},\tilde{z})\}$ still can be transformed to the dimensional version $\{\phi(R,z),\Phi(R,z)\}$ by $\lambda\approx6\times10^{-7}$ under the transformation of Eq.~(\ref{eq:symmetry}).

\begin{figure}[]
\begin{center}
\includegraphics[width= 9.4cm]{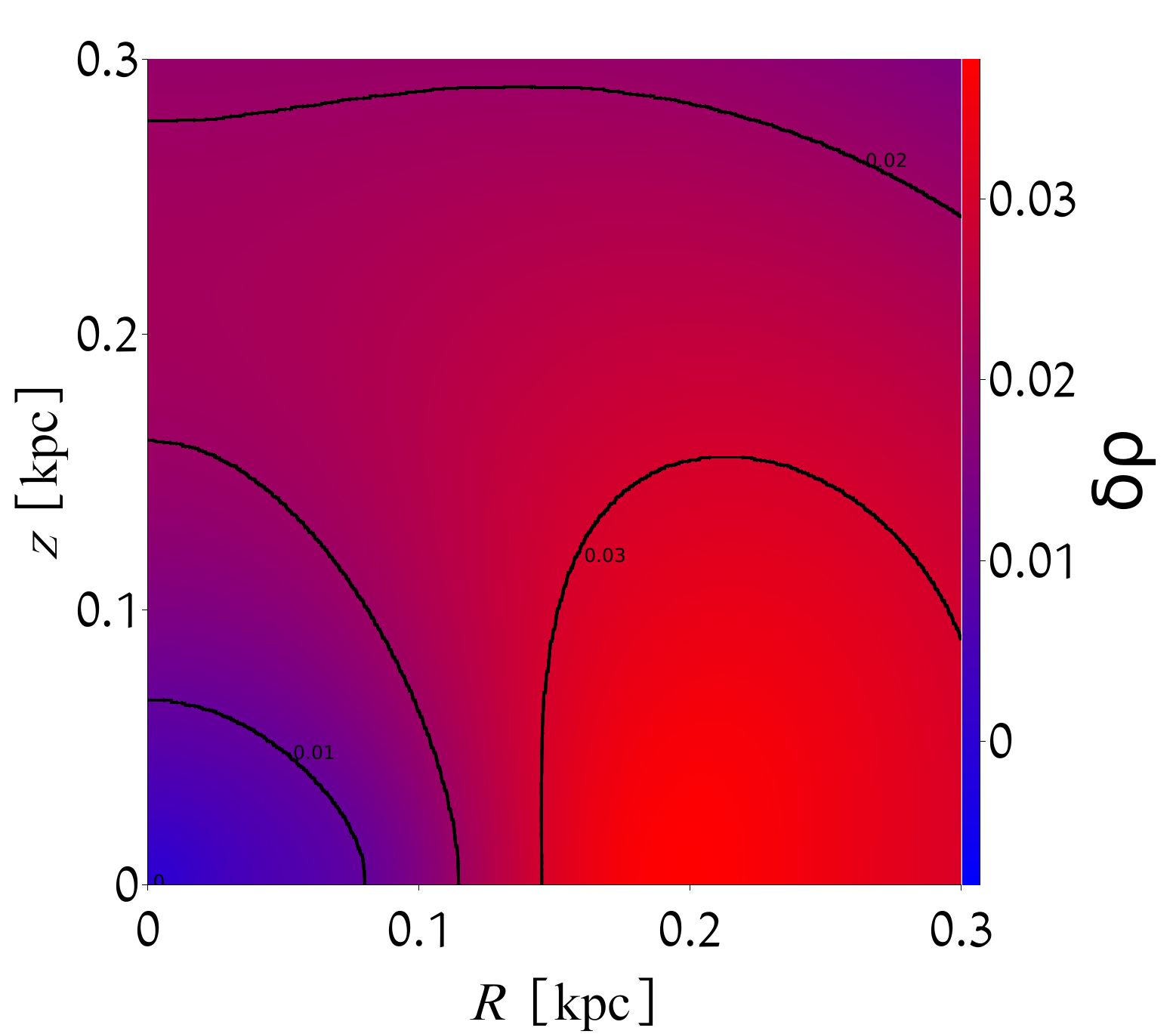}
\end{center}
\caption{Distribution of fractional error of $\delta\rho(R,z)$.}
\label{fig:drho}
\end{figure}

\begin{figure}[]
\begin{center}
\includegraphics[width= 9.4cm]{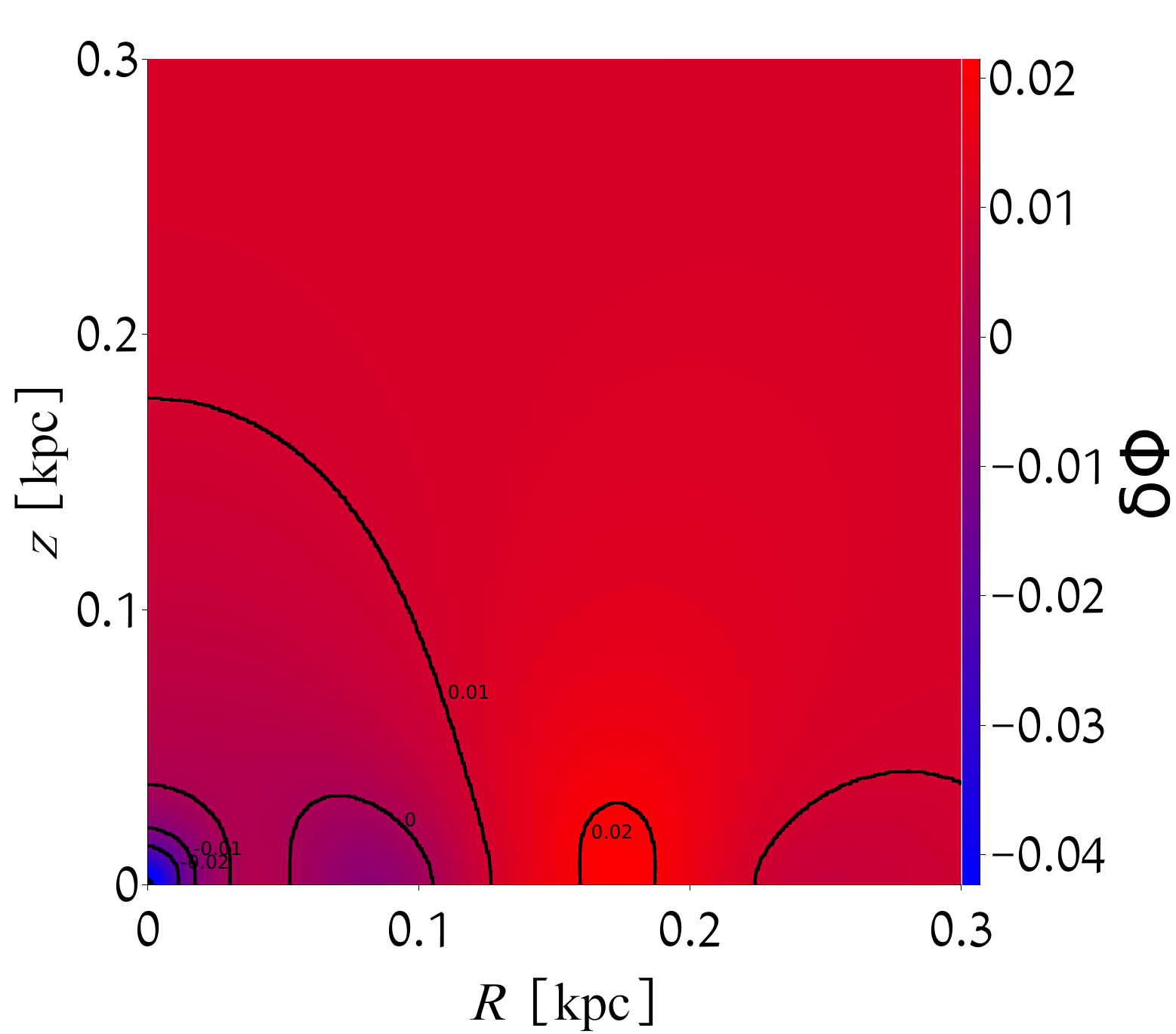}
\end{center}
\caption{Distribution of fractional error of $\delta\Phi(R,z)$.}
\label{fig:dPhi}
\end{figure}

If not focusing on the Milky Way, given a different fixed baryon background $\tilde{\rho}_{\rm b}$, there should be a different solution of $\{\tilde{\phi}(\tilde{R},\tilde{z}),\tilde{\Phi}(\tilde{R},\tilde{z})\}$ for Eq.~(\ref{eq:sp1}). Similarly, to find the solutions of Eq.~(\ref{eq:sp2}) for the Milky Way, we also need to impose some conditions on $\tilde{\rho}_{\rm b,AE}$. Here, we first fix the integral of $\tilde{\rho}_{\rm b,AE}(\tilde{R},\tilde{z})$ as one of Fig.~\ref{fig:rhoRz}, $\int\tilde{\rho}_{\rm b,AE}=\int\tilde{\rho}_{\rm b}=\tilde{M}_{\rm b}=1.03$. Since different baryon density profiles can also lead to the same value of $\tilde{M}_{\rm b}$, we further impose another condition at one space point, where $\tilde{\Phi}(\tilde{R}=0,\tilde{z}=0)=-2.75395$, $\tilde{\phi}^2(\tilde{R}=0,\tilde{z}=0)=1$ and $\tilde{\rho}_{\rm b,AE}(\tilde{R}=0,\tilde{z}=0)=21551.2$.
Finally, unlike solving Eq.~(\ref{eq:sp1}) with iteration of $\lambda$~\cite{Tan:2024dne}, Eq.~(\ref{eq:sp2}) under above two conditions can be solved directly in cylindrical coordinates by following the Appendix~A of~\cite{Bar:2019bqz}. In Fig.~\ref{fig:drho} and Fig.~\ref{fig:dPhi}, we compare the dimensional counterparts of $\{\tilde{\phi}(\tilde{R},\tilde{z}),\tilde{\Phi}(\tilde{R},\tilde{z})\}$ for Eq.~(\ref{eq:sp1}) and Eq.~(\ref{eq:sp2}), where
\begin{align}
\delta\rho(R,z)&=\frac{\phi^2_{\rm Eq13}(R,z)-\phi^2_{\rm Eq3}(R,z)}{\phi^2_{\rm Eq3}(R,z)},\\
\delta\Phi(R,z)&=\frac{\Phi_{\rm Eq13}(R,z)-\Phi_{\rm Eq3}(R,z)}{\Phi_{\rm Eq3}(R,z)}.
\end{align}
We find both of $|\delta\rho(R,z)|$ and $|\delta\Phi(R,z)|$ are $\lesssim0.04$, which is consistent with the accuracy of our baryon \Mov{AE} $<4\%$, as shown in Fig.~\ref{fig:acc}.

\section{Summary and discussion}
\label{sec:sum} 
In this paper, we first solve the SP system with a fixed empirical baryon density profile background in cylindrical coordinates by an iteration of $\lambda$. For $m=10^{-22}\rm{eV}/c^2$, $\lambda\approx6\times10^{-7}$ rescales the dimensionless $\tilde{\rho}_{\rm b}$ and $\tilde{M}$ to the baryon density profile and the FDM soliton mass of the Milky Way simultaneously. Therefore, $\lambda\approx6\times10^{-7}$ was used to map the scaled dimensionless variables to the corresponding physical dimensional ones and vice versa. Then we make the obtained stationary solutions mock data and construct an analytical emulator (with $\%{\rm Acc}<4\%$) for the baryon \Mov{density distribution} inside the FDM soliton by GA, using a machine learning technique. Finally, we complement a model validation by comparison between the stationary solutions of an enlarged SP system with the emulator $\tilde{\rho}_{\rm b,AE}$ as the third equation under certain conditions and ones of the original SP system. We find the fractional errors of two sets of solutions are $\lesssim0.04$ and are consistent with the accuracy of baryon \Mov{AE} $<4\%$.

What about its validation for the nonstationary evolutions of FDM solitons, especially those evolutions totally departing from the stationary solutions, such as FDM soliton collisions.
Since the mock data used in this paper is extracted from the stationary solutions of the SP system with a fixed empirical baryon density profile background, our baryon \Mov{AE} inside the FDM soliton should also be valid \Mov{and can serve as a kind of baryon EoM} for the perturbation, the tidal disruption or deformation of FDM solitons conservatively.
However, for some dramatic evolutions, our baryon \Mov{AE} probably does not perform correctly. 

\vspace{5mm}
\noindent {\bf Acknowledgments}
We acknowledge the use of HPC Cluster of Tianhe II in National Supercomputing Center in Guangzhou. Ke Wang is supported by grant from the National Key Research and Development Program of China (grant No. 2021YFC2203003). Jianbo Lu is supported by the National Natural Science Foundation of China (12175095), and supported by LiaoNing Revitalization Talents Program (XLYC2007047). Man Ho Chan is supported by the grant from the Research Grants Council of the Hong Kong Special Administrative Region, China (Project No. EdUHK 18300324).




\begin{thebibliography}{99}
\frenchspacing
\bibitem{Rubin:1982kyu}
V.~C.~Rubin, W.~K.~Ford, Jr., N.~Thonnard and D.~Burstein,
``Rotational properties of 23 SB galaxies,''
Astrophys. J. \textbf{261}, 439 (1982)

\bibitem{Davis:1985rj}
M.~Davis, G.~Efstathiou, C.~S.~Frenk and S.~D.~M.~White,
``The Evolution of Large Scale Structure in a Universe Dominated by Cold Dark Matter,''
Astrophys. J. \textbf{292}, 371-394 (1985)

\bibitem{Clowe:2006eq}
D.~Clowe, M.~Bradac, A.~H.~Gonzalez, M.~Markevitch, S.~W.~Randall, C.~Jones and D.~Zaritsky,
``A direct empirical proof of the existence of dark matter,''
Astrophys. J. Lett. \textbf{648}, L109-L113 (2006)
[arXiv:astro-ph/0608407 [astro-ph]].

\bibitem{Hu:2000ke}
W.~Hu, R.~Barkana and A.~Gruzinov,
``Cold and fuzzy dark matter,''
Phys. Rev. Lett. \textbf{85}, 1158-1161 (2000)
[arXiv:astro-ph/0003365 [astro-ph]].

\bibitem{Guzman:2004wj}
F.~S.~Guzman and L.~A.~Urena-Lopez,
``Evolution of the Schrodinger-Newton system for a selfgravitating scalar field,''
Phys. Rev. D \textbf{69}, 124033 (2004)
[arXiv:gr-qc/0404014 [gr-qc]].

\bibitem{Bernal:2006ci}
A.~Bernal and F.~Siddhartha Guzman,
``Scalar Field Dark Matter: Head-on interaction between two structures,''
Phys. Rev. D \textbf{74}, 103002 (2006)
[arXiv:astro-ph/0610682 [astro-ph]].

\bibitem{Paredes:2015wga}
A.~Paredes and H.~Michinel,
``Interference of Dark Matter Solitons and Galactic Offsets,''
Phys. Dark Univ. \textbf{12}, 50-55 (2016)
[arXiv:1512.05121 [astro-ph.CO]].

\bibitem{Edwards:2018ccc}
F.~Edwards, E.~Kendall, S.~Hotchkiss and R.~Easther,
``PyUltraLight: A Pseudo-Spectral Solver for Ultralight Dark Matter Dynamics,''
JCAP \textbf{10}, 027 (2018)
[arXiv:1807.04037 [astro-ph.CO]].

\bibitem{Munive-Villa:2022nsr}
E.~Munive-Villa, J.~N.~Lopez-Sanchez, A.~A.~Avilez-Lopez and F.~S.~Guzman,
``Solving the Schr\"odinger-Poisson system using the coordinate adaptive moving mesh method,''
Phys. Rev. D \textbf{105}, no.8, 083521 (2022)
[arXiv:2203.10234 [gr-qc]].

\bibitem{Tan:2024spp}
C.~Tan, J.~K.~Bin and K.~Wang,
``Gravitational Waves from Post-Collision of Fuzzy Dark Matter Solitons,''
[arXiv:2412.19262 [astro-ph.CO]].

\bibitem{Davies:2019wgi}
E.~Y.~Davies and P.~Mocz,
``Fuzzy Dark Matter Soliton Cores around Supermassive Black Holes,''
Mon. Not. Roy. Astron. Soc. \textbf{492}, no.4, 5721-5729 (2020)
[arXiv:1908.04790 [astro-ph.GA]].

\bibitem{Bar:2019bqz}
N.~Bar, K.~Blum, J.~Eby and R.~Sato,
``Ultralight dark matter in disk galaxies,''
Phys. Rev. D \textbf{99}, no.10, 103020 (2019)
[arXiv:1903.03402 [astro-ph.CO]].

\bibitem{Tan:2024dne}
C.~Tan, M.~L.~Delliou and K.~Wang,
``Diversity of fuzzy dark matter solitons,''
Phys. Rev. D \textbf{112}, no.6, 063021 (2025)
[arXiv:2411.16114 [hep-ph]].

\bibitem{Launhardt:2002tx}
R.~Launhardt, R.~Zylka and P.~G.~Mezger,
``The nuclear bulge of the galaxy. 3. Large scale physical characteristics of stars and interstellar matter,''
Astron. Astrophys. \textbf{384}, 112-139 (2002)
[arXiv:astro-ph/0201294 [astro-ph]].

\bibitem{Chan:2019ukj}
M.~H.~Chan,
``A universal constant for dark matter-baryon interplay,''
Sci. Rep. \textbf{9}, no.1, 3570 (2019)
[arXiv:1902.03786 [astro-ph.CO]].

\bibitem{Chan:2022fzq}
M.~H.~Chan,
``Two mysterious universal dark matter{\textendash}baryon relations in galaxies and galaxy clusters,''
Phys. Dark Univ. \textbf{38}, 101142 (2022)
[arXiv:2212.01018 [astro-ph.GA]].

\bibitem{McGaugh:2004aw}
S.~S.~McGaugh,
``The Mass discrepancy - acceleration relation: Disk mass and the dark matter distribution,''
Astrophys. J. \textbf{609}, 652-666 (2004)
[arXiv:astro-ph/0403610 [astro-ph]].

\bibitem{McGaugh:2016leg}
S.~McGaugh, F.~Lelli and J.~Schombert,
``Radial Acceleration Relation in Rotationally Supported Galaxies,''
Phys. Rev. Lett. \textbf{117}, no.20, 201101 (2016)
[arXiv:1609.05917 [astro-ph.GA]].

\bibitem{Collett:2019hrr}
T.~Collett, F.~Montanari and S.~Rasanen,
``Model-Independent Determination of $H_0$ and $\Omega_{K0}$ from Strong Lensing and Type Ia Supernovae,''
Phys. Rev. Lett. \textbf{123}, no.23, 231101 (2019)
[arXiv:1905.09781 [astro-ph.CO]].

\bibitem{Huang:2017huh}
Q.~G.~Huang and K.~Wang,
``Effect of the Early Reionization on the Cosmic Microwave Background and Cosmological Parameter Estimates,''
JCAP \textbf{07}, 042 (2017)
[arXiv:1704.08495 [astro-ph.CO]].

\bibitem{Shafieloo:2012ht}
A.~Shafieloo, A.~G.~Kim and E.~V.~Linder,
``Gaussian Process Cosmography,''
Phys. Rev. D \textbf{85}, 123530 (2012)
[arXiv:1204.2272 [astro-ph.CO]].

\bibitem{Seikel:2012uu}
M.~Seikel, C.~Clarkson and M.~Smith,
``Reconstruction of dark energy and expansion dynamics using Gaussian processes,''
JCAP \textbf{06}, 036 (2012)
[arXiv:1204.2832 [astro-ph.CO]].

\bibitem{Yang:2024rsy}
Z.~F.~Yang, D.~W.~Yao, M.~Le Delliou and K.~Wang,
``Model-independent test of the cosmic anisotropy with inverse distance ladder,''
Eur. Phys. J. C \textbf{85}, no.3, 339 (2025)
[arXiv:2407.19278 [astro-ph.CO]].

\bibitem{Wang:2019vxv}
G.~J.~Wang, X.~J.~Ma, S.~Y.~Li and J.~Q.~Xia,
``Reconstructing Functions and Estimating Parameters with Artificial Neural Networks: A Test with a Hubble Parameter and SNe Ia,''
Astrophys. J. Suppl. \textbf{246}, no.1, 13 (2020)
[arXiv:1910.03636 [astro-ph.CO]].

\bibitem{Nesseris:2012tt}
S.~Nesseris and J.~Garcia-Bellido,
``A new perspective on Dark Energy modeling via Genetic Algorithms,''
JCAP \textbf{11}, 033 (2012)
[arXiv:1205.0364 [astro-ph.CO]].

\bibitem{Arjona:2019fwb}
R.~Arjona and S.~Nesseris,
``What can Machine Learning tell us about the background expansion of the Universe?,''
Phys. Rev. D \textbf{101}, no.12, 123525 (2020)
[arXiv:1910.01529 [astro-ph.CO]].

\bibitem{Orjuela-Quintana:2022nnq}
J.~B.~Orjuela-Quintana, S.~Nesseris and W.~Cardona,
``Using machine learning to compress the matter transfer function T(k),''
Phys. Rev. D \textbf{107}, no.8, 08 (2023)
[arXiv:2211.06393 [astro-ph.CO]].

\bibitem{Orjuela-Quintana:2023uqb}
J.~B.~Orjuela-Quintana, S.~Nesseris and D.~Sapone,
``Machine learning unveils the linear matter power spectrum of modified gravity,''
Phys. Rev. D \textbf{109}, no.6, 063511 (2024)
[arXiv:2307.03643 [astro-ph.CO]].

\bibitem{Orjuela-Quintana:2024hha}
J.~B.~Orjuela-Quintana, D.~Sapone and S.~Nesseris,
``Fully Interpretable Emulator for the Linear Matter Power Spectrum from Physics-Informed Machine Learning,''
[arXiv:2407.16640 [astro-ph.CO]].

\bibitem{Bartlett:2023cyr}
D.~J.~Bartlett, L.~Kammerer, G.~Kronberger, H.~Desmond, P.~G.~Ferreira, B.~D.~Wandelt, B.~Burlacu, D.~Alonso and M.~Zennaro,
``A precise symbolic emulator of the linear matter power spectrum,''
Astron. Astrophys. \textbf{686}, A209 (2024)
doi:10.1051/0004-6361/202348811
[arXiv:2311.15865 [astro-ph.CO]].

\bibitem{Setyawati:2019xzw}
Y.~Setyawati, M.~P{\"u}rrer and F.~Ohme,
``Regression methods in waveform modeling: a comparative study,''
Class. Quant. Grav. \textbf{37}, no.7, 075012 (2020)
[arXiv:1909.10986 [astro-ph.IM]].

\bibitem{Khan:2020fso}
S.~Khan and R.~Green,
``Gravitational-wave surrogate models powered by artificial neural networks,''
Phys. Rev. D \textbf{103}, no.6, 064015 (2021)
[arXiv:2008.12932 [gr-qc]].

\bibitem{Thomas:2022rmc}
L.~M.~Thomas, G.~Pratten and P.~Schmidt,
``Accelerating multimodal gravitational waveforms from precessing compact binaries with artificial neural networks,''
Phys. Rev. D \textbf{106}, no.10, 104029 (2022)
[arXiv:2205.14066 [gr-qc]].

\bibitem{Wang:2019ubx}
W.~Wang, J.~Han, M.~Cautun, Z.~Li and M.~N.~Ishigaki,
``The mass of our Milky Way,''
Sci. China Phys. Mech. Astron. \textbf{63}, no.10, 109801 (2020)
[arXiv:1912.02599 [astro-ph.GA]].

\bibitem{Schive:2014hza}
H.~Y.~Schive, M.~H.~Liao, T.~P.~Woo, S.~K.~Wong, T.~Chiueh, T.~Broadhurst and W.~Y.~P.~Hwang,
``Understanding the Core-Halo Relation of Quantum Wave Dark Matter from 3D Simulations,''
Phys. Rev. Lett. \textbf{113}, no.26, 261302 (2014)
[arXiv:1407.7762 [astro-ph.GA]].

\end{thebibliography}
\end{document}